# Universal band center model for the HER activity of non-metal site

Ruixin Xu[1], Shiqian Cao[1], Tingting Bo[1], Yanyu Liu[2,*], Wei Zhou[1,*]

[1]*Department of Applied Physics, Tianjin Key Laboratory of Low Dimensional Materials Physics and Preparing Technology, School of Science, Tianjin University, Tianjin 300072, P. R. China.*
[2]*College of Physics and Materials Science, Tianjin Normal University, Tianjin, 300387, P. R. China.*

**ABSTRACT**. In this work, the hydrogen evolution reaction activities of non-metal sites in the 144-transition metal dichalcogenides with the stoichiometry of $MX_2$ are investigated using the first principles calculations. The trained machine learning model demonstrates that the $p_z$ band center, bandgap, and period effect are the key factors influencing the HER activity of $MX_2$. Furthermore, it also reveals that the observed non-scaling law between the $p_z$ band center and HER activity in the semiconductor-like materials originates from the bandgap-induced downshift of bonding and antibonding states. In addition to the bandgap, the intrinsic $p$ orbital energy level of the non-metal atoms also contributes to the periodic variation of $p_z$ band center. Extended calculations indicate that the descriptor is equally applicable to the other catalysts, suggesting its universality in predicting the HER activity of non-metal sites.

## I. INTRODUCTION.

Recently, surface reaction [1-3] continues to be a hot topic in the fields of physics and chemistry. Especially, a deep understanding of localized surface interaction is essential when catalytic processes occur at the surface [4]. How to interpret the surface phenomena and uncover its intrinsic mechanism interaction is a crucial challenge to the rational design and exploration of potential catalysts [5-7].

The $d$ band center ($\varepsilon_d$) model [8, 9], proposed by Nørskov *et al.*, plays a crucial role in the area of surface catalytic process. Over the past decades, the model has achieved great success in characterizing the interactions between metals and adsorbates [10-13]. In general, $d$ band center is considered as the descriptor for predicting the adsorption strength between adsorbate and catalyst, where the reactive site with low $\varepsilon_d$ will result in a weaker adsorption capacity. However, $d$ band center model is based on metal reactive sites with continuous $d$ orbitals. In the case of metals without $d$ orbitals or nonmetals, other analogous approaches, *e.g.*, $p$ band center, should be used [14].

Back in 2011, Morgan *et al.* [15] demonstrated that the experimental measurements of area specific resistance and oxygen surface exchange of solid oxide fuel cell cathode perovskites are tightly linked to the $p$ band center ($\varepsilon_p$) of oxygen and vacancy formation energy calculated by the density functional theory (DFT). The calculated reaction energy/activation potential barrier correlation comparisons with the O-$p$ band center are also markedly better than the $d$ band center. To date, $\varepsilon_p$ is increasingly proved to be the effective descriptor for revealing the catalytic activity of non-metal reactive sites [16, 17]. However, with the discovery of new non-metal catalysts, the non-scaling law between the $p$ band center and catalytic activity [18-22]. This hampers the deep understanding of the surface catalytic mechanism. Consequently, it is urgent to reveal the origin of the divergence of $p$ band center model.

Transition metal dichalcogenides (TMDs) [23-27] are the materials with rich structural polymorphism and novel physical properties. This allows them to provide more active sites in hydrogen evolution reaction (HER), promote charge transport, and further optimize catalytic performance. In this work, as the prototype materials of TMDs, the origin of HER activity is investigated using the first principles calculations. The bandgap induced divergence in the traditional $p$ band center is revealed. The modified band center model considering the bandgap and atomic orbital energy level effect is proposed. And it proved as a universal descriptor for predicting the HER activity of TMDS and other catalysts. It also offers a new physical insight for comprehending the surface catalytic mechanism of non-metal active sites.

## II. METHODS

The spin-polarized DFT calculations are carried out via Vienna ab initio simulation package (VASP) [28, 29] with the projector augmented wave (PAW) method [30]. The generalized gradient approximation (GGA) with Perdew-Burke-Ernzerhof (PBE) functional [31] are employed for the electron exchange-correlation interactions. The cutoff energy is set to be 400 eV in this work. The Brillouin zones are sampled through the Monkhorst-Pack 3×3×1 k-point grids [32], and DFT-D2 correction method [33] is employed to improve the description of the van der Waals (vdW) interactions. The energy and force convergence tolerance are $10^{-5}$ eV and 0.01 eV/Å, respectively. A vacuum space with a width of 20 Å is constructed along the z-axis to diminish the interactions between adjacent layers. To obtain the bonding information, the projected crystal orbital Hamilton population ($p$COHP) analysis is performed by the LOBSTER program [34].

## III. RESULTS

*HER activity of MX$_2$*—Transition metal dichalcogenides (TMDs) with the stoichiometry of MX$_2$ (M = transition metals, X = chalcogens) exhibit a variety of attractive properties and span a relatively large polymorphic phase space [35]. Among them, the monolayer TMDs have two basic structural units: the trigonal prismatic phase and the octahedral phase, which belong to the D$_{3h}$ and D$_{3d}$ groups, respectively. In the following discussions, they are referred to as 1H-MX$_2$ and 1T-MX$_2$, respectively. Previous studies have demonstrated that MX$_2$ with the post-transition metal atoms (IB and IIB) typically struggles to form stable layered structures [23, 36]. Therefore, as shown in Fig. 1(a), we constructed a comprehensive library of MX$_2$ models covering 24 different metals, three sulfur elements (S, Se, Te), and two phases (1H and 1T) for extensive first principles calculations. That is to say, the total number of MX$_2$ models considered here is 144. As a reference, the atomic structures of the 1H and 1T phases are shown in Fig. 1(b).

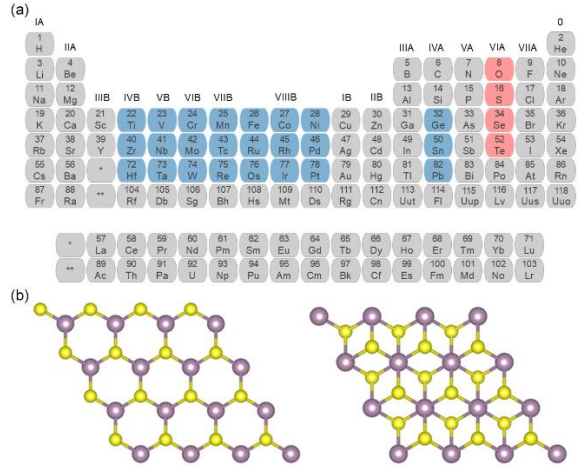

FIG 1. (a) Periodic table of elements. The highlighted blue and pink colors indicate the metal and nonmetal atoms considered in this work, respectively. (b) The crystal structures of 1H-MX$_2$ and 1T-MX$_2$, respectively. (c) In-plane Brillouin zone of the hexagonal unit cell.

Since the basal plane of monolayer MX$_2$ only exposes non-metal (X) atoms, X atom is considered as the H adsorption site here. For all MX$_2$ models, the calculated hydrogen adsorption free energy ($\Delta G_{H*}$) ranges from −0.27 to 2.43 eV, as shown in Fig. 2(a). In general, the absolute value of $\Delta G_{H*}$ is closer to 0 eV [37], the hydrogen evolution reaction (HER) performance is higher. Larger $\Delta G_{H*}$ results in weaker H adsorption strength and vice versa. Most of the calculated $\Delta G_{H*}$ of MX$_2$ are greater than 0 eV, implying the suitable strategy to enhancing their adsorption ability for H atom is needed to improve the HER activity of MX$_2$. To quantitatively describe the HER activity, the relationship between $\Delta G_{H*}$ and the exchange current density ($i_0$) [37] is shown in Fig. 2(b). $i_0$ is calculated by the following equation:

$$i_0 = -ek_0 \frac{1}{1+\exp(|\Delta G_{H*}|/k_B T)} \quad (1)$$

where $e$, $k_0$, $k_B$, and $T$ are the elementary charge, reaction rate constant, Boltzmann constant, and temperature, respectively. Notably, 1H-TiS$_2$ with the $\Delta G_{H*}$ of −0.01 eV is very close to the peak of the volcano plot and exhibits the best HER activity, which is comparable or even higher than the conventional Pt-based catalysts (−0.09 eV) [38].

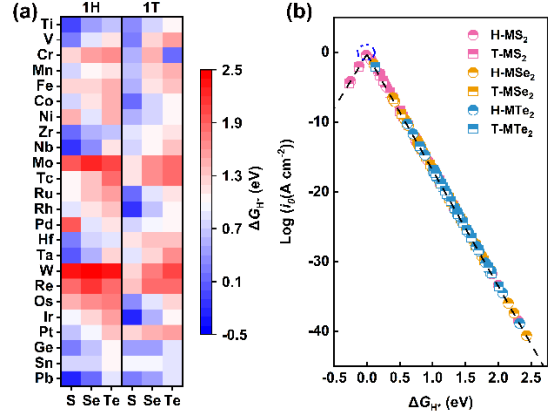

FIG 2. (a) Gibbs free energy diagram for HER on 1H-MX$_2$ and 1T-MX$_2$. (b) Plot of the exchange current density *vs.* $\Delta G_{H*}$.

Furthermore, after structural optimization, it is evident that the bond length between X and H atoms shows a periodic increase with the atomic number of X atom (Table S1). Meanwhile, corresponding to the same X atom, linear relationship between the bond length of X-H and $\Delta G_{H*}$ is shown in Fig. S1. It indicates that with the increase in bond length, the value of $\Delta G_{H*}$ becomes more positive, which is accompanied by the corresponding weakening of H adsorption. Nevertheless, for the different X atoms, the $\Delta G_{H*}$ of MX$_2$ shows periodic variations with the bond length, which suggests that the periodic variation of the surface X atoms could result in the difference of the basal properties without affecting H adsorption strength. Bader charge analysis also reveals that the electron gain/loss of the H atom shows a periodic variation with the atomic number of X element (Table S2, Fig. S2), mainly related to the difference in the electronegativity of the X and H atoms (S: 2.58, Se: 2.55, Te: 2.1, and H: 2.2). This is also in agreement with the changes of bond lengths between X and H atoms mentioned above.

*Origin of HER activity*—In order to explore the physical origin of the HER activity in TMDs, their electronic structures have been systematically analyzed. In fact, when H atom is adsorbed on the $MX_2$, the H-$s$ and X-$p_z$ orbitals would undergo hybridization interactions because of their extended z-axis adsorption direction, which could be clearly seen from the orbital overlap of the H-$s$ and X-$p_z$ orbitals in the projected density of states (PDOS) [Fig. 3(a, b)]. Based on the modern molecular orbital (MO) theory, atomic orbitals undergo linear combination to form molecular orbitals during the atomic bonding process, including bonding molecular orbitals with decreasing energy levels and antibonding molecular orbitals with increasing energy levels. Electrons are populated in molecular orbitals with energy levels in ascending order. Bonding orbital electrons contributes to strengthen the bonding interaction, whereas antibonding orbital electrons act to weaken it. Unoccupied orbitals do not play a role in affecting the bonding strength. In general, the more the energy levels of the bonding orbitals fall and the antibonding orbitals rise, which forms the more stable molecular orbitals and enhances the bonding strength.

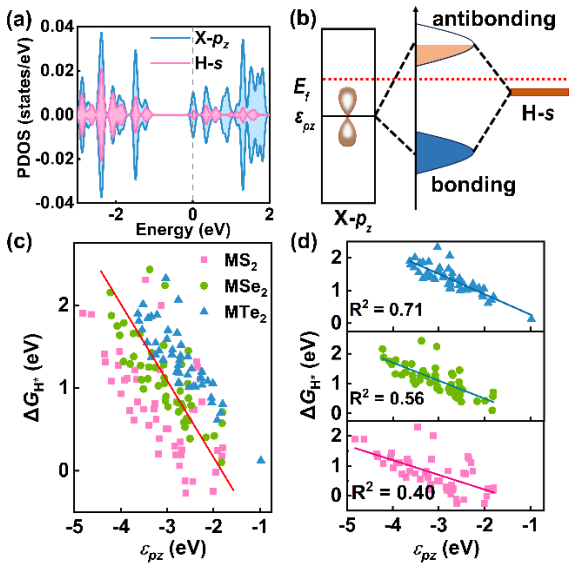

FIG 3. (a) PDOS of H-$s$ and X-$p_z$ orbitals. (b) Schematic diagram of the interaction between H-$s$ and X-$p_z$ orbitals. (c) Plot of the $\varepsilon_{pz}$ vs. $\Delta G_{H^*}$ for all $MX_2$. (d) Plot of the $\varepsilon_{pz}$ vs. $\Delta G_{H^*}$ for all $MS_2$, $MSe_2$, and $MTe_2$, respectively.

In addition, the binding strength of the adsorbate in the potential-determining step (PDS) is closely related to the properties of the valence orbitals of the reactive site that determine the filling of the antibonding state near the Fermi level [39-41]. Thus, $p_z$ band centers of the reactive site in the $MX_2$ are calculated, and it is obviously that there is a highly diffuse inverse trend between the X-$p_z$ band center and $\Delta G_{H^*}$ [Fig. 3(c)]. Although a clear inverse trend could be seen for the different X atoms separately, the determination coefficients ($R^2$) are relatively low, with the unsatisfactory values of 0.40, 0.56, and 0.71, respectively [Fig. 3(d)]. It can be found that the relatively smaller value of $R^2$ could be attributed primarily to individual discrete data points, for which electronic structure analysis reveals that all these models exhibit semiconductor-like material with bandgap. The detailed $MX_2$ and their corresponding bandgap values are listed on Table S3. It is worth noting that for the different X atoms, it seems to be a translational relationship between the trend lines of the X-$p_z$ band center and $\Delta G_{H^*}$, which may be related to the periodic differences of different X atoms mentioned above. In order to quantitatively depict the bonding strength, an integration of $p$COHP (the projected crystal orbital Hamilton populations) is performed to obtain the integrated COHP (ICOHP). Unfortunately, there is no linear relationship between the calculated ICOHP and $\Delta G_{H^*}$ (Fig. S3), and a similar translational relationship as above could be seen from most of the systems with the specific X atoms. These both imply that there could exist other determinants for the adsorption strength of H atom.

*Training ML models for HER of $MX_2$*—In order to reveal the underlying structure-activity relationship of $MX_2$, the scikit-learn library [42] is utilized to construct the Random Forest Regression (RFR) model [43], in which 21 initial metrics are taken into account as input variables to identify the key factors influencing the HER performance of $MX_2$. The features consist of four main parts, i) the atomic features of M and X atoms, including the atomic number ($N$), atomic mass ($M$), and covalent radius ($R_{cov}$); ii) the electronic structure features of M and X atoms, including the electronegativity ($X$), first ionization energy ($E_I$), affinity energy ($E_A$), $d$ electrons number of M atom ($\theta_d$), and valence electrons number of M atom ($\theta_{val}$); iii) geometrical structure features of materials, including the bond length between M and X atoms ($d_{M-X}$), and the bond length between X and H atoms ($d_{X-H}$); iv) electronic structure features of materials, including the $d$ band center of M ($\varepsilon_d$), $p_z$ band center of X ($\varepsilon_{pz}$), bandgap ($\varepsilon_{gap}$), net charge of X atom ($Q_X$), and net charge of H atom ($Q_H$).

Subsequently, high feature-feature correlations are removed by determining the Pearson correlation coefficient (PCC) among paired features (|PCC|>0.70). As shown in Fig. 4a, 11 features exhibiting high independence, $N_M$, $R_{cov-M}$, $R_{cov-X}$, $\theta_{val-M}$, $X_M$, $E_{I-X}$, $E_{A-M}$, $\varepsilon_{d-M}$, $\varepsilon_{pz-X}$, $\varepsilon_{gap}$, and $Q_X$, serve as the input set for predicting the HER performance. It is worth noting that in the strongly correlated feature pairs, $Q_H$, $d_{X-H}$, and $d_{M-X}$ are all correlated with the periodic atomic structure changes of X atom ($N_X$, $M_X$, $R_{cov-X}$), which is also consistent with the above statement. In order to mitigate the inaccuracies stemming from the random dataset division, five-fold cross-validation strategy is employed. The final results are shown in Fig. 4(b), where RFR shows a high $R^2$ (0.98) as well as a small RMSE (0.087), demonstrating the suitability and adaptability of the RFR model in predicting the HER activity of $MX_2$. Apart from that, as shown in Fig. 4(c), the importance of features in descending order is: $\varepsilon_{p_z\text{-}X}$ (38.39%) > $\varepsilon_{gap}$ (14.08%) > $E_{I\text{-}X}$ (10.51%) > $R_{cov\text{-}X}$ (10.47%) > $\varepsilon_{d\text{-}M}$ (7.88%) > $R_{cov\text{-}M}$ (5.59%) > $Q_X$ (4.79%) > $X_M$ (2.77%) > $N_M$ (2.35%) > $\theta_{val\text{-}M}$ (1.68%) > $E_{A\text{-}M}$ (1.49%). The cumulative importance of the first four features, $\varepsilon_{p_z\text{-}X}$, $\varepsilon_{gap}$, $X_X$, and $R_{cov\text{-}X}$, is more than 73%, which suggests that the HER activities of $MX_2$ are chiefly dependent on the electronic properties of the X atoms and the material. Surprisingly, the bandgap of materials contributes 14.08% of the significance, which could be related to the anomalous phenomenon of the band center applied to semiconductor-like materials. Furthermore, since the first ionization energy and covalent radius of the same X atom are identical, meaning that there are just three pairs of identical features out of the 144 data points, yet their importance amounts to almost 21%. It also corresponds to the effect of the period effect of the X atoms on the architecture mentioned above.

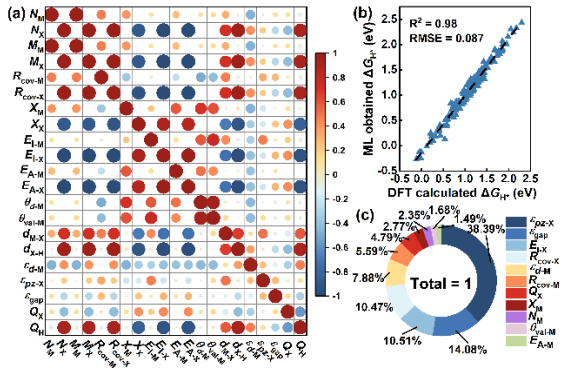

FIG 4. (a) Pearson correlation coefficient plots of 21 initial variables. (b) Plot of the $\Delta G_{H*}$ computed by DFT *vs.* $\Delta G_{H*}$ obtained by RFR model for all $MX_2$. (c) The relative importance rankings of the 11 input features.

*The modified band center model*—Obviously, in order to reveal the intrinsic mechanism of the above phenomenon, we need to re-check the formula of $p_z$ band center as follows,

$$\varepsilon_{p_z} = \frac{\int_{-\infty}^{0} \varepsilon \rho_{p_z}(\varepsilon) d\varepsilon}{\int_{-\infty}^{0} \rho_{p_z}(\varepsilon) d\varepsilon} \qquad (2)$$

where $\varepsilon$ denotes the energy level and $\rho_{p_z}(\varepsilon)$ indicates the DOS distribution of X-$p_z$ orbital. Essentially, the formula primarily computes the energy levels of $p_z$ orbital using a weighted-average approach, where the DOS distribution serves as the weighting factor. Obviously, it only takes into account the energy level information of the electronic orbitals in the pre-adsorption system, yet the energy levels of the bonding and antibonding orbitals are severely neglected.

As shown in Fig. 5(a-f), the changes in the electronic structure of the $MX_2$ systems before and after H adsorption are analyzed using 1H-$MoS_2$ with a bandgap value of 1.30 eV and 1T-$VS_2$ without bandgap as examples, which have similar $p_z$ band centers and vastly different $\Delta G_{H*}$. It can be found that 1H-$MoS_2$ exhibits an upward shift of Fermi level after H adsorption. However, due to the existence of bandgap, there will be a huge range of empty bands in the valence band, which is different from the case in 1T-$VS_2$. It leads to a downward shift of the bonding and antibonding states of 1H-$MoS_2$ after H adsorption. According to the MO theory, the downshift of the antibonding state formed by the hybridization of the H-$s$ orbital and the S-$3p_z$ orbital in 1H-$MoS_2$ will result in a weaker H adsorption and a larger $\Delta G_{H*}$. Both of these suggest that the formation of bandgap may give the similar band centers but with the widely differing $\Delta G_{H*}$, and the difference is dependent on the value of bandgap. Hence, we consider that the DOS around the Fermi level before H adsorption in the formula of $p_z$ band center may corresponds to the position lower than the Fermi level with the value of bandgap after H adsorption, and the DOS that originally is around the conduction band minimum (CBM) may correspond to the position at the lower-part of Fermi level after H adsorption. Here, for the sake of model simplicity, the value of the specific deviation is not considered. Considering $p_z$ band center to calculate the weights for the new DOS ($\varepsilon + \varepsilon_{gap}$), and after the coordinate transformation, the original band center would be written as,

$$\varepsilon_{p_z}^{re} = \frac{\int_{-\infty}^{0} \varepsilon \rho'_{p_z}(\varepsilon) d\varepsilon}{\int_{-\infty}^{0} \rho'_{p_z}(\varepsilon) d\varepsilon} \qquad (3)$$

where $\rho'_{p_z}(\varepsilon) = \rho_{p_z}(\varepsilon + \varepsilon_{gap})$. Therefore,

$$\begin{aligned}\varepsilon_{p_z}^{re} &= \frac{\int_{-\infty}^{0} \varepsilon \rho_{p_z}(\varepsilon + \varepsilon_{gap}) d\varepsilon}{\int_{-\infty}^{0} \rho_{p_z}(\varepsilon + \varepsilon_{gap}) d\varepsilon} \\ &= \frac{\int_{-\infty}^{0} (\varepsilon' - \varepsilon_{gap}) \rho_{p_z}(\varepsilon') d(\varepsilon' - \varepsilon_{gap})}{\int_{-\infty}^{0} \rho_{p_z}(\varepsilon') d(\varepsilon' - \varepsilon_{gap})} \\ &= \frac{\int_{-\infty}^{\varepsilon_{gap}} \varepsilon' \rho_{p_z}(\varepsilon') d\varepsilon' - \varepsilon_{gap} \int_{-\infty}^{\varepsilon_{gap}} \rho_{p_z}(\varepsilon') d\varepsilon'}{\int_{-\infty}^{\varepsilon_{gap}} \rho_{p_z}(\varepsilon') d\varepsilon'} \\ &= \frac{\int_{-\infty}^{\varepsilon_{gap}} \varepsilon' \rho_{p_z}(\varepsilon') d\varepsilon'}{\int_{-\infty}^{\varepsilon_{gap}} \rho_{p_z}(\varepsilon') d\varepsilon'} - \varepsilon_{gap}\end{aligned} \qquad (4)$$

and due to the DOS is 0 between 0 eV and $\varepsilon_{gap}$, therefore,

$$\varepsilon_{p_z}^{re} = \frac{\int_{-\infty}^{0} \varepsilon \rho_{p_z}(\varepsilon) d\varepsilon}{\int_{-\infty}^{0} \rho_{p_z}(\varepsilon) d\varepsilon} - \varepsilon_{gap} \quad (5)$$

$$= \varepsilon_{p_z} - \varepsilon_{gap}$$

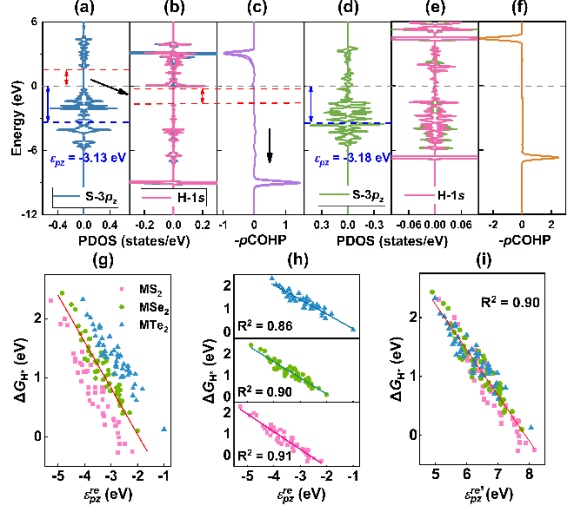

FIG 5. PDOS of X-$p_z$ and H-$s$ orbitals for 1H-MoS$_2$ (a) before and (b) after H adsorption. (c) -$p$COHP for H adsorption on 1H-MoS$_2$. PDOS of X-$p_z$ and H-$s$ orbitals for 1T-VS$_2$ (d) before and (e) after H adsorption. (f) -$p$COHP for H adsorption on 1T-VS$_2$. (g) Plot of the $\varepsilon_{p_z}^{re}$ vs. $\Delta G_{H^*}$ for all MX$_2$. (h) Plot of the $\varepsilon_{p_z}^{re}$ vs. $\Delta G_{H^*}$ for all MS$_2$, MSe$_2$, and MTe$_2$, respectively. (i) Plot of the $\varepsilon_{p_z}^{re'}$ vs. $\Delta G_{H^*}$ for all MX$_2$.

It is an ingenious equation which could be simply described as the difference between $p_z$ band center and band gap. However, as shown in Fig. 5(g,h), it could still be found that it only applies to the MX$_2$ with the same X atom. For the different X atoms, the period effect of X element still induced differences in the $p_z$ band center. Therefore, according to the predicted results of the RFR model, the first ionization energy of X atoms ($E_{I-X}$), which corresponds to the energy required to remove the outermost $p$ orbital electron of X atoms and directly reflects the absolute value of the $p$ orbital energy level, is considered to eliminate the bias caused by the X atoms themselves. Based on the NIST atomic spectra database [44], the first ionization energy of S, Se, and Te atoms are −10.36, −9.75, and −9.06 eV, respectively. Hence, the final modified band center model and predicted equation for $\Delta G_{H^*}$ would be written as,

$$\varepsilon_{p_z}^{re'} = \varepsilon_{p_z} - \varepsilon_{gap} + E_{I-X} \quad (6)$$

$$\Delta G_{H^*} = -0.77 \varepsilon_{p_z}^{re'} + 6.10 \quad (7)$$

It is astonishing that the HER activity of all MX$_2$ could be well depicted by $\varepsilon_{p_z}^{re'}$ with $R^2$ of 0.90, as shown in Fig. 5 (i). In fact, the modified band center model could be equally extended to other common two-dimensional (2D) layered metal chalcogenides such as InS, InSe, InTe, GeS, GeSe, SnS, SnSe, Bi$_2$Se$_3$, Bi$_2$Te$_3$, and so on [45, 46]. The corresponding structures and space groups of the materials are shown in Fig. S4a. The modified X-$p_z$ band center and $\Delta G_{H^*}$ are also calculated with non-metal atom as the H adsorption site. As shown in Fig. S4b, the data points of the above-mentioned materials are well integrated into all the data points of MX$_2$, which suggests the universality of the model in predicting the HER activity of non-metal sites.

## IV. CONCLUSIONS

In summary, the first principles calculations are employed to analyze the anomaly of the $p_z$ band center in transition metal dichalcogenides with the stoichiometry of MX$_2$. It is found that the band gap of the semiconductor-like materials leads to the significant deviations from the traditional $p_z$ band center model, which hampers the accurate prediction of hydrogen evolution reaction activity. Meanwhile, the overall offset of $p_z$ band center for different non-metal X atoms originates from the differences in their $p_z$ orbital energy levels. The derived modified model not only elucidates the abnormal behavior of the band center model in MX$_2$ but also achieves higher accuracy in predicting the HER activity of other catalysts. This work provides novel physical insight into the catalytic processes occurring on the non-metal sites.

## ACKNOWLEDGMENTS
This work was supported by the National Natural Science Foundation of China (51972227).

## APPENDIX
Supplemental data associated with this article can be found in the **Supplemental Material**.

*Corresponding authors.
yyliu@tjnu.edu.cn
weizhou@tju.edu.cn

[1] X. Kang et al., Nat. Commun. 14, 3607 (2023).
[2] P. Li, Y. Jiang, Y. Hu, Y. Men, Y. Liu, W. Cai, and S. Chen, Nat. Catal. 5, 900 (2022).
[3] Z. Y. Wu et al., Nat. Mater. 22, 100 (2023).
[4] M. Konsolakis, Catalysts 6 (2016).
[5] Z. Niu, Z. Lu, Z. Qiao, S. Wang, X. Cao, X. Chen, J. Yun, L. Zheng, and D. Cao, Adv. Mater. 36, e2310690 (2024).
[6] H. Xu, D. Cheng, D. Cao, and X. C. Zeng, Nat. Catal. 1, 339 (2018).

[7] H. Xu, D. Cheng, D. Cao, and X. C. Zeng, Nat. Catal. 7, 207 (2024).
[8] B. Hammer and J. K. Nørskov, Nature 376, 238 (1995).
[9] B. Hammer and J. K. Nørskov, Surf. Sci. 343, 211 (1995).
[10] B. H. A. Ruban, P. Stoltze, H.L. Skriver, J.K. Nørskov, J. Mol. Catal. 115, 421 (1997).
[11] S. Huang et al., Appl. Catal. B-Environ. 317 (2022).
[12] J. K. N. Line S. Byskov, Bjerne S. Clausen, Henrik Topsøe, J. Catal. 187, 109 (1999).
[13] Z. Xiao, P. Sun, Z. Qiao, K. Qiao, H. Xu, S. Wang, and D. Cao, Chem. Eng. J. 446 (2022).
[14] X. Wang, G. Zhang, L. Yang, E. Sharman, and J. Jiang, WIREs Comput. Mol. Sci. 8 (2018).
[15] Y.-L. Lee, J. Kleis, J. Rossmeisl, Y. Shao-Horn, and D. Morgan, Energ. Environ. Sci. 4 (2011).
[16] B. Ren et al., Nat. Commun. 13, 2486 (2022).
[17] B. Sun et al., Small 19, e2207461 (2023).
[18] R. Jiang, Z. Qiao, H. Xu, and D. Cao, Chinese J. Catal. 48, 224 (2023).
[19] R. Jiang, Z. Qiao, H. Xu, and D. Cao, Nanoscale 15, 16775 (2023).
[20] R. Xu, T. Bo, S. Cao, N. Mu, Y. Liu, M. Chen, and W. Zhou, J. Mater. Chem. A 10, 21315 (2022).
[21] K. Liu, J. Fu, Y. Lin, T. Luo, G. Ni, H. Li, Z. Lin, and M. Liu, Nat. Commun. 13, 2075 (2022).
[22] R. Xu, T. Lin, S. Cao, T. Bo, Y. Liu, and W. Zhou, J. Energy Chem. 99, 636 (2024).
[23] T. Lu, Y. Wang, G. Cai, H. Jia, X. Liu, C. Zhang, S. Meng, and M. Liu, Mater. Futures 2 (2023).
[24] Z. Lu, G. P. Neupane, G. Jia, H. Zhao, D. Qi, Y. Du, Y. Lu, and Z. Yin, Adv. Funct. Mater. 30 (2020).
[25] M. Wu, Y. Xiao, Y. Zeng, Y. Zhou, X. Zeng, L. Zhang, and W. Liao, InfoMat 3, 362 (2020).
[26] M. Xu, T. Liang, M. Shi, and H. Chen, Chem. Rev. 113, 3766 (2013).
[27] S. Joseph, J. Mohan, S. Lakshmy, S. Thomas, B. Chakraborty, S. Thomas, and N. Kalarikkal, Mater. Chem. Phys. 297 (2023).
[28] G. Kresse and J. Hafner, Phys. Rev. B 47, 558 (1993).
[29] Kresse and Furthmuller, Phys. Rev. B 54, 11169 (1996).
[30] P. E. Blochl, Phys. Rev. B 50, 17953 (1994).
[31] J. P. Perdew, J. A. Chevary, S. H. Vosko, K. A. Jackson, M. R. Pederson, D. J. Singh, and C. Fiolhais, Phys. Rev. B 46, 6671 (1992).
[32] H. J. Monkhorst and J. D. Pack, Phys. Rev. B 13, 5188 (1976).
[33] S. Grimme, J. Comput. Chem. 27, 1787 (2006).
[34] S. Maintz, V. L. Deringer, A. L. Tchougreeff, and R. Dronskowski, J. Comput. Chem. 37, 1030 (2016).
[35] M. Chhowalla, H. S. Shin, G. Eda, L. J. Li, K. P. Loh, and H. Zhang, Nat. Chem. 5, 263 (2013).
[36] X. Wu, M. Sun, H. Yu, B. Huang, and Z. L. Wang, Nano Energy 115 (2023).
[37] J. K. Nørskov, T. Bligaard, A. Logadottir, J. R. Kitchin, J. G. Chen, S. Pandelov, and U. Stimming, J. Electrochem. Soc. 152 (2005).
[38] P. G. M. Berit Hinnemann, Jacob Bonde, Kristina P. Jørgensen, Jane H. Nielsen, Sebastian Horch, Ib Chorkendorff, and Jens K. Nørskov*, J. Am. Chem. Soc. 127, 5308 (2005).
[39] J.-H. Lee, D. Yim, J. H. Park, C. H. Lee, J.-M. Ju, S. U. Lee, and J.-H. Kim, J. Mater. Chem. A 8, 13490 (2020).
[40] J. H. Park, C. H. Lee, J. M. Ju, J. H. Lee, J. Seol, S. U. Lee, and J. H. Kim, Adv. Funct. Mater. 31 (2021).
[41] N. K. Wagh, S. S. Shinde, C. H. Lee, J.-Y. Jung, D.-H. Kim, S.-H. Kim, C. Lin, S. U. Lee, and J.-H. Lee, Appl. Catal. B-Environ. 268 (2020).
[42] J. R. Kitchin, Nat. Catal. 1, 230 (2018).
[43] R. Diaz-Uriarte and S. Alvarez de Andres, BMC Bioinf. 7, 3 (2006).
[44] Kramida, A., Ralchenko, Yu., Reader, J., and NIST ASD Team (2024). NIST Atomic Spectra Database (ver. 5.12).
[45] J. Yin et al., Adv. Sci. 7, 1903070 (2020).
[46] J. Zhou et al., Nat. Mater. 22, 450 (2022).